# Structural and Magnetic Characterizations of $Co_2FeGa/SiO_2$ Nanoparticles Prepared via Chemical Route

Priyanka and Rajendra S. Dhaka[*]

*Novel Materials and Interface Physics Laboratory, Department of Physics,
Indian Institute of Technology Delhi (IIT/D), Hauz Khas, New Delhi-110016, India*

[*]Corresponding author: rsdhaka@physics.iitd.ac.in

**Abstract.** We report the synthesis of $Co_2FeGa/SiO_2$ nanoparticles by sol-gel method and characterization using x-ray diffraction (XRD), transmission electron microscopy (TEM) and magnetic measurements. The Rietveld refinements of XRD data with space group Fm-3m clearly show the formation of A2 disorder single phase and the lattice constant is found to be 5.738 Å. The energy-dispersive x-ray spectroscopy (EDX) confirm the elemental composition close the desired values. The value of coercivity is found to be around 283 Oe and 126 Oe, measured at 10 K and 300 K, respectively. We observed the saturation magnetization significantly lower than expected from Slater-Pauling rule. This decrease in the magnetic moment might be due to the presence of amorphous $SiO_2$ during the synthesis process. A large content of small size $SiO_2$ particles along with $Co_2FeGa$ nanoparticles are also found in TEM study.

## INTRODUCTION

The magnetic nanoparticles, because of their small size and large surface to volume ratio, exhibit novel applications such as data storage devices, magnetic sensors, drug and gene delivery, DNA sequencing, magnetic resonance imaging (MRI) etc. In recent years, Heusler alloy nanoparticles have also attracted great attention due to their ferromagnetic nature [1]. Especially, Co-based Heusler nanopartciles are of particular ineterst because of their high Cuire temperature well above the room temperature. For the first time, Basit *et al.* synthesized and investigated ternary $Co_2FeGa$ nanoparticles by simple chemical method [2]. The short range ordering in these nanoaprticles can influence their structural and magnetic properties [3]. A new class of shape memory effect in chemically synthesized nanoaprtciles has been discovered by Wang et al. [4]. With very high Curie temperature (~1174 K), the size dependent structural and magnetic properties of CoNiAl nanoparticles have been studied where increasing the particle size increases the Curie temperature ($T_c$) and magnetization. In a recent work, a new method via chemical route using simple reduction method has been reported for nanoparticle fabrication [5-6]. To best of our knowledge the synthesis of these nanoparticles and study their detailed magnetic and structural properties are largely unexplored in the literature. It is important to understand the effect of the nanoparticle size on their physical and magnetic properties. Therefore, we studied the influence of annealing temperature ($T_A$) and different hydrogen environment during annealing on their morphological, magnetic and structural properties. We optimize the synthesis procedure of ternary $Co_2FeGa$ nanoparticles in $SiO_2$ matrix and report their structural/magnetic characterizations.

## EXPERIMENTAL

The $Co_2FeGa$ nanoparticles were synthesized by chemical route, where 0.48 mmol (0.194gm) $Fe(NO_3)_3 \cdot 9H_2O$, 0.49 mmol (0.117gm) $CoCl_2 \cdot 6H_2O$ and 0.32 mmol (0.128gm) $Ga(NO_3)_3 \cdot xH_2O$ ($x$=8 here) were prepared in 50 ml methanol. All the precursors were used as purchased without any further purification. For decomposition and reduction of Co, Fe and Ga precursor the solution was kept for sonication of 5 min. Further 1g fumed silica was added to this solution and mixed in ultra-sonicator bath for 1h. After that when the resulting solution turned to yellowish solution, to evaporate the solvent completely, the solution was stirred continuously using a magnetic stirrer at room temperature overnight until we get a sticky gel form. Then, the obtained gel has been dried in air at 90°C for 2h to achieve complete dryness. In order to get fine homogeneous dense powder, this solid was gently ground to powder for few minutes and annealed at 600-1000°C for 5h in 10% or 5% $H_2$ balanced Ar flow environment. The final powder was allowed to cool up to room temperature and collected for characterization.

We use x-ray diffraction (XRD) with Cu K$\alpha$ ($\lambda$ = 1.5406 Å) radiation for structural study and analyzed the XRD data by Rietveld refinement using FullProf package where the background was fitted using linear interpolation between the data points. The compositions have been confirmed by energy dispersive x-ray spectroscopy (EDX) measurements. For TEM measurement the sample was deposited from methanol suspension on copper grid. TEM is performed in microscope JEOL JEM-1400 Plus with 120 KV accelerating voltage. The magnetic measurements are performed using vibrating sample magnetometer (VSM) mode in a physical properties measurements system (PPMS EVERCOOL- II) from Quantum design.

## Results and Discussion

In Fig. 1, we compare the XRD data of $Co_2FeGa$ nanoparticles annealed from 600 to 1000°C temperature range in different $H_2$ flow environment. At $T_A$=600°C, there is huge background and significant hump at 2$\theta$ ~20° as well as the 220 peak intensity is very small with Co impurity, as seen in Fig.1 (a). We can see that higher annealing temperature eliminates the Co impurity and improve the crystallinity. For example, for $T_A$=700°C to 1000°C, the 220 peak is much sharper and highest intensity. We observed single phase with all the peaks (220), (400), and (422) corresponds to the A2 disorder phase in Heusler alloys.

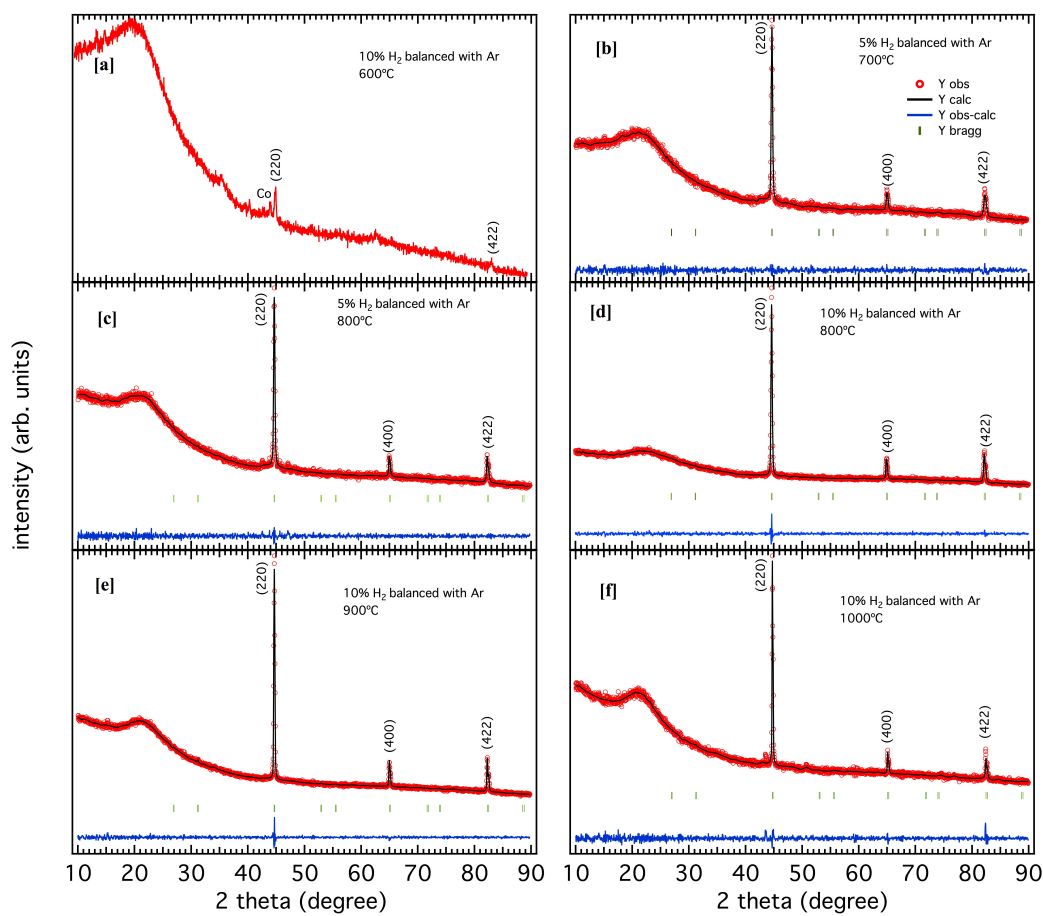

**FIGURE 1.** The powder XRD data (open circles) and Rietveld refinement (black line) of $Co_2FeGa/SiO_2$ nanoparticles samples with difference profile (blue line) and Bragg peak positions (vertical bars) annealed at (a) 600°C ($x$=10%), (b) 700°C ($x$=5%), (c) 800°C ($x$=5%), (d) 800°C ($x$=10%), (e) 900°C ($x$=10%) and (f) 1000°C ($x$=10%) under $x$%$H_2$ plus balance Ar flow.

Note that the structure is thermally stable in the temperature range of 700-1000°C, as shown in Fig. 1. The Rietveld refinement of powder XRD data with the space group Fm-3m (225) is shown in the Figs.1 (b-f). The lattice constants and FRHM are found to be about $a = b = c$ = 5.738 Å and 0.27°, respectively for sample annealed at 800°C in 5% $H_2$ flow and there is no significant change with different annealing temperatures. The crystallite size is estimated about 31 nm by using Scherrer formula for the samples annealed at $T_A$=700°C to 1000°C. In order to find the purity and elemental composition, $Co_2FeGa$ nanoparticles (annealed at 800°C) have been analyzed using EDX (not shown), which confirms the stoichiometry close to the desired values. However, we observed about 60% of total signal from silica only, which is excluded in the composition calculation of $Co_2FeGa$ nanoparticles. The hump at around 2θ=20° in all XRD pattern is the signature of amorphous $SiO_2$, and the absence of (200) and (111) Bragg reflection provides the evidence of existence of A2 disorder. However, additional probes such as neutron diffraction or extended x-ray absorption fine structure (EXAFS) are required to detect the existence of B2 and A2 disorder. Further, we use the sample annealed at 800 in 5% $H_2$ for other studies such as TEM and magnetic measurements.

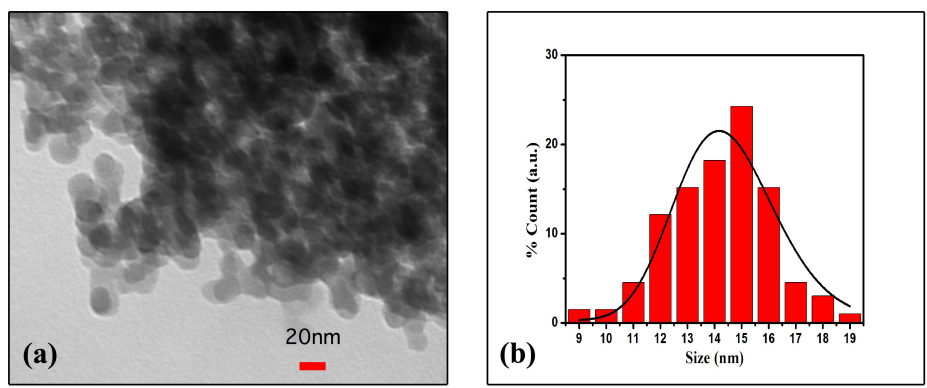

**FIGURE 2.** (a) TEM image and (b) particle size histogram of $Co_2FeGa/SiO_2$ nanoparticles annealed at 800°C in 5% $H_2$ flow.

The particle size and size distribution of the $Co_2FeGa$ nanoparticles are studied using TEM. Figs. 2 (a, b) show the TEM image and the extracted particle size distribution histogram of $Co_2FeGa$ nanoparticles, respectively. The TEM image shows that these particles are spherical in shape and the particle size distribution is not wide in range. We plotted the histogram in Fig. 2(b) by estimating the size of large number of individual particles from Fig. 2(a). The histogram is fitted with the lognormal fit (solid line) in the range between 9 and 19 nm, which gives the average particle size of 14 nm with standard deviation of 0.2 nm. However, the particle size calculated from TEM is smaller that the crystallite size extracted from XRD (31 nm). This indicates that the large content of small particles in the TEM image might be due to amorphous $SiO_2$ particles along with the $Co_2FeGa$ nanoparticles.

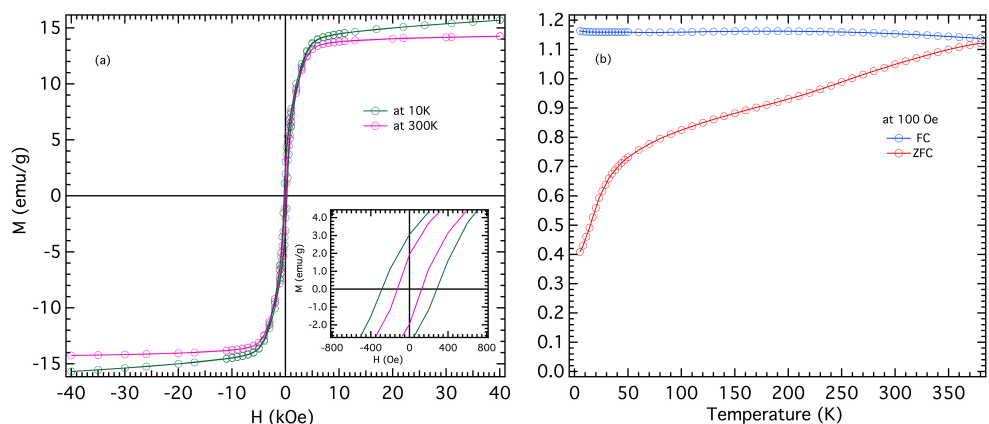

**FIGURE 3.** (a) Magnetization (M-H) curve at 300 K (pink) and 10 K (green), (b) magnetization v/s temperature (FC, ZFC) curve at 100 Oe for $Co_2FeGa$ nanoparticles annealed at 800°C in 5% $H_2$ flow.

Fig. 3(a) show the isothermal magnetization curves of $Co_2FeGa$ nanoparticles, measured at two different temperatures. The value of coercivity is found to be significantly larger ($H_C$ = 283 Oe) at 10 K as compare to the value ($H_C$ = 126 Oe) measured at 300 K. However, the saturation magnetization values are comparable *i.e.* 14.4 and 13.8 emu/g measured at 10 K and 300 K, respectively. Note that half-metallic full Heusler alloys follow a Slater-Pauling behavior, *i.e.*, the total spin magnetic moment per formula unit, $M_t$, in $\mu_B$ scales with the total number of valence electrons, $Z_t$, following the rule: $M_t=Z_t-24$. As per Slater-Pauling rule, the total moment of $Co_2FeGa$ sample is expected to be about 5 $\mu_B$/*f.u.* (~115 emu/g). Note that the observed moment is much lower than the expected from the Slater-Pauling rule, which could be due to the presence of large amount of amorphous $SiO_2$. For $Fe_3O_4$ nanoparticles it is reported that $SiO_2$ is used to control the particle size; however, increasing the amount of $SiO_2$ decrease the magnetization [7]. The zero-field-cooled (ZFC) and field-cooled (FC) magnetization measured in the temperature range of 5-380 K at 100 Oe are shown in Fig. 3(b). The FC magnetization is almost independent of temperature in this range; however, the ZFC magnetization decreases and show the change in the slope at ~50K. Below 50 K, the ZFC curve shows rapid decrease up to 0.4 emu/g at 5 K. The ZFC and FC data indicate that the blocking ($T_B$) and Curie ($T_c$) temperatures are higher than 380 K. Due to the small size distribution of nanoparticles, it may show a sharp peak at $T_B$ in the ZFC curve. The higher temperature magnetic susceptibility measurements are useful to find the blocking and Curie temperatures.

In conclusion, we successfully prepared ternary $Co_2FeGa$ nanoparticles via wet chemical method. Here, we optimize the synthesis procedure of these nanoparticles in the $SiO_2$ matrix and present their structural /magnetic characterization. The XRD data confirm the crystalline single phase of the samples annealed above 700°C. The lattice constant value of about 5.738 Å is obtained by Rietveld refinement (with space group Fm-3m), which found to be close to the theoretical predicted value. The presence of large amount of amorphous silica can be seen by EDX and XRD. The obtained nanoparticles are spherical in shape with average particle size of 14 nm. The significantly lower magnetization can also be attribute to the presence of large amount of amorphous $SiO_2$. It would be interesting to see the effect by varying the amount of $SiO_2$ during the synthesis process. Also, the magnetic susceptibility measurements at high temperature are required to understand the nature and find the Curie temperature.

## ACKNOWLEDGMENTS

Priyanka acknowledges the MHRD, India for fellowship. Authors thank IIT Delhi for various experimental facilities: XRD, PPMS EVERCOOL-II at Physics department, TEM at CRF, and EDX, AGM at NRF. We also thank the physics department of IIT Delhi for excellent support. RSD gratefully acknowledges the financial support from BRNS through DAE Young Scientist Research Award project sanction No. 34/20/12/2015/BRNS.